# The morphological identification of the rapidly evolving population of faint galaxies

Karl Glazebrook[1], Richard Ellis[1], Basilio Santiago[1] and Richard Griffiths[2]

[1] *Institute of Astronomy, Madingley Road, Cambridge CB3 0HA*
[2] *Department of Physics and Astronomy, Johns Hopkins University, 3400 North Charles St, Baltimore MD21218, USA*

**ABSTRACT**
The excess numbers of blue galaxies at faint magnitudes are a long-standing cosmological puzzle. We present new number-magnitude counts as a function of galactic morphology from the first deep fields of the Cycle 4 Hubble Space Telescope *Medium Deep Survey* project. From a sample of 301 galaxies we define counts for elliptical, spiral and irregular/peculiar galaxies to $I = 22$. We find two principal results. Firstly the elliptical and spiral galaxy counts both follow the predictions of high-normalisation no-evolution models at all magnitudes, indicating that regular Hubble types evolve only slowly to $z \sim 0.5$. Secondly we find that irregular/peculiar galaxies, including multiple-peaked, possibly merging, objects, have a very steep number-magnitude relation and greatly exceed predictions based on proportions in local surveys. These systems make up half the total counts by $I = 22$ and imply the rapidly-evolving component of the faint galaxy population has been identified.

## INTRODUCTION

Counting galaxies as a function of magnitude is a traditional yet active tool in extragalactic astronomy. Originally a cosmological test (Hubble 1926), it is now recognised to be more sensitive to galaxy evolution (Brown & Tinsley 1974) for which evidence is accumulating. In particular, counts of galaxies in the optical passbands have revealed a dramatic excess of 'faint blue galaxies' (for a review see Ellis 1990). Extensive redshift surveys have been undertaken of these objects selected in $B$ (Broadhurst *et al.* 1988, Colless *et al.* 1990, Glazebrook *et al.* 1995A, Ellis *et al.* 1994), $I$ (Lilly 1993, Tresse *et al.* 1994) and $K$ bands (Glazebrook *et al.* 1995B, Cowie *et al.* 1994) which have been used to construct the respective luminosity functions as they evolve with redshift. At short wavelengths, where the evolution implied by the counts is strongest, there appears to be an increase in the space-density of galaxies at luminosity $M_B \sim -20$ (we use $H_0 = 100\,\mathrm{km\,s^{-1}\,Mpc^{-1}}$), over $0 < z < 0.5$ (Ellis *et al.* 1994). At longer $I$ and $K$ wavelengths, the excess appears to occupy fainter portions of the luminosity function (Glazebrook *et al.* 1995B, Cowie *et al.* 1994).

There are still some major uncertainties of interpretation. Firstly there is the fundamental problem of $\phi^*$, the local normalisation of the Schechter luminosity function, which remains uncertain to within a factor of two. Clearly the higher the local galaxy density, the less evolution has to be invoked. Local surveys have given a low value; Efstathiou *et al.* (1988) and Loveday *et al.* (1992A) both find $\phi^* \simeq 0.015h^3\,\mathrm{Mpc^{-3}}$ based on redshift surveys to $B = 17$ where the median redshift is $z \sim 0.05$ and the effects of large-scale structure should be negligible.

Adopting this normalisation leads to a factor of two excess of galaxies as bright as $B = 19$, where the median redshift is only 0.1, increasing to a factor of four by $B = 24$ and $z = 0.5$. This has led Maddox *et al.* (1990) to argue for extreme evolution at low-redshift. On the other hand Metcalfe *et al.* (1991, 1994A, 1994B) and Koo *et al.* (1993) have argued that the bright data is not representative: possibly being subject to selection effects, large-scale structure or problems with the local galaxy luminosity function. They prefer to normalise at $B = 19$, equivalent to $\phi^* \simeq 0.03h^3\,\mathrm{Mpc^{-3}}$, which still leaves an excess at $B = 24$ of $\times 2$ which requires evolutionary explanations; it is not possible to explain the whole of it via local effects. Regardless of $\phi^*$ the slope of the counts is steeper than the no-evolution model over $15 < B < 19$ and this points to some anomalous behaviour in the local data.

Secondly the excess at $B = 24$ is now established as being caused by an increase by $z = 0.5$ in the density of objects with $M_B = -20$ (Glazebrook *et al.* 1994, Ellis *et al.* 1994), i.e. of similar luminosity to local giant ellipticals and spirals. What has caused this population to disappear by $z = 0$? Proposed explanations include: (i) extreme differential luminosity evolution in which dwarf galaxies are preferentially promoted from the faint end of the local luminosity function (Broadhurst *et al.* 1988) at $z \sim 0.5$; (ii) merger models in which the extra population is merged into the local population by $z = 0$ (Rocca-Volmerange & Guideroni 1990, Broadhurst *et al.* 1992 (BEG)) and (iii) 'exploding dwarf' models (Cowie *et al.* 1991) in which the blue galaxies are a new population of small starbursting systems: the starburst is powerful and dominated by high-mass stars; the remaining gas is expelled by supernova driven winds preventing further star-formation and turning them into dark systems by $z = 0$ (Babul & Rees 1992).

Ideally we would like to directly identify the population of objects contributing to the excess and track its evolution with redshift. This has been attempted using the strength of the [OII] line as a star-formation indicator (BEG) and the population of such objects in the $B$ luminosity function has been found to evolve rapidly (Ellis *et al.* 1994). However [OII] emission may be a transient feature and it is more sound to examine structural morphology — this requires sub-arcsecond resolution to see kpc scale structures at $z \sim 0.5$. Limited work has been done from the ground in $0.5-0.8''$ seeing; this shows that a high fraction of the star-forming galaxies to $B = 22.5$ have multiple cores (Colless *et al.* 1993).

## OBSERVATIONS & DISCUSSION

The new data we describe were taken as part of the *Medium Deep Survey*, a key project designed to maximize HST science by taking WFPC2 or FOC frames in parallel, while another instrument observes a primary target (Griffiths *et al.* 1994A). Thus a large sample of random fields results with exposures of several thousand seconds and reaching depths of $I > 20$ (note unless stated we use $I = $ F814W). Our particular sample consists of all 13 deep fields with total exposures $\geq 4000$ secs observed with the WFPC2 camera since the start of Cycle 4, when the spherical aberration was corrected. The data were reduced in a standard way: pipeline bias/dark/flat corrections were applied and the data aligned and coadded using a cosmic ray filter. Cosmic ray removal is especially critical due to the high incidence in HST data; we use the 'crreject' algorithm as described in Griffiths *et al.* which is found to work well for WFPC2 images and our sample excludes fields with only 1 or 2 individual exposures and galaxies on the PC chip which has a different pixel scale. The resulting total area is $51.3 \, \mathrm{arcmin}^2$.

Object selection is another critical issue for HST data, as a galaxy which produces a single image from the ground can when resolved, have a complex structure. For example it would be incorrect to split spiral arms into multiple objects and it is unclear whether merging objects should be counted as one or two objects. Because we wish to have a sample which closely corresponds to the 'objects' seen in previous ground-based counts we smoothed the HST data with a $1''$ FWHM Gaussian profile and then select objects above a fixed isophote of typically $25.5 \, \mathrm{mags/arcsec}^2$, as if it was ground-based data. We then return to the unsmoothed data and classify each 'object' by detailed visual examination.

Blue galaxies are seen in abundance at a mean redshift of $\sim 0.5$ where the WFPC2 pixel scale of $0.1''$ corresponds to $\sim 0.4 h^{-1}$ kpc. Since the PSF FWHM $\simeq 0.04''$ the data is under sampled and the effective PSF is close to a delta function in one pixel. Objects of galactic size are typically $\gtrsim 10$ resolution elements across; the basic star-galaxy separation is straight forward. Our total galaxy counts are shown in Figure 1(a) and compared with those from the ground (Tyson 1988, Lilly *et al.* 1991, Driver 1994). We use a small colour correction of $I(\mathrm{F814W}) = I(\mathrm{Cousins}) - 0.07$ which represents the transformation of Harris *et al.* 1991 for a mean galaxy colour over $19 < I < 22$ of $(V - I)_c = 1.5$ (Driver 1994, our data); we independently measured the likely $I_{814} - I_c$ colours with our galaxy spectral templates

(described below) and got similar numbers. Figure 1(a) shows our counts are complete to $I = 23$ and agree well with the previous work.

For our classified sample we consider all galaxies to $I = 22$ (301 total in all fields) irrespective of size; this allows us to obtain a clean magnitude limited sample which can be modeled in detail. At $I = 22$ ($\sim \equiv B = 24, K = 19$) the excess over the no-evolution prediction is $\times 2$–$4$ (depending on local normalisation) and the median redshift for representative samples is $\simeq 0.5$ (Lilly 1993). Obviously the more detailed a classification is attempted the more unreliable it becomes; differences between close types (e.g. Sa/Sb) become subjective. In this paper we are interested in only a basic classification of the fundamental morphological types, viz:

i) Elliptical/S0 galaxies. These show smooth profiles with no spiral features or disks. We see a range of sizes including a class of compact objects, contributing $\simeq 20\%$ of the objects in each magnitude bin.

ii) Spiral galaxies. We require these to show an extended disk with evident spiral arms, or to be clearly edge-on disks. Note in some cases the arms are seen to be distorted or show strong starbursts, but if the underlying structure is clearly a classic spiral we assign this class.

iii) Irregular/peculiar galaxies. This category contains sources with peculiar or disturbed irregular structure. Many have low surface brightnesses compared to the previous two classes. We also include objects that appear to be merging or closely-interacting (at sub-$1''$ resolution).

This first-order classification is fairly unambiguous and allows robust conclusions to be drawn. An extensive atlas of typical images of such MDS galaxies of various morphologies has already been presented in Griffiths *et al.* 1994B. In this paper we now consider in detail the number-magnitude counts as a function of morphology, plotted in Figure 1(b–d), to $I < 22$ and compare with no-evolution expectations.

For our local luminosity function we use a determination of Schechter parameters ($M^*, \alpha$) for early and late type galaxies taken from Loveday *et al.* (1992A) based on 1658 APM galaxies to $b_J < 17$. Zucca, Pozzetti & Zamorani (1994) have reanalysed this dataset under different assumptions for classification errors at faint magnitudes and have suggested that the early type galaxies might have a flat faint end to their luminosity function rather than a turnover. We will consider this possibility. For a more detailed analysis we also use the morphological mix to $b_J < 16.7$ given by Shanks *et al.* 1984, adjusting the weight of each type in the luminosity function to produce the observed flux-limited mix. Our K-corrections are based on spectral templates of Rocca-Volmerange & Guiderdoni (1988) which faithfully reproduce local galaxy UV-optical colours.

The resulting models are plotted in Figure 1 for both normalisations: $\phi^* = 0.015 \, h^3 \, \mathrm{Mpc}^{-3}$ and $\phi^* = 0.03 \, h^3 \, \mathrm{Mpc}^{-3}$. For the ellipticals we also show the effect of a flat faint end to the luminosity function. Several points are of interest. Our total counts reproduce the excess noted in earlier work. However our counts for spiral and ellipticals show good agreement with the no-evolution prediction if we accept the higher normalisation. Moreover the slopes match the data indicating that any evolution would have to be mild. The effect of a flat faint end slope in the ellipti-cal luminosity function affects only the very faint data and cannot yet be tested thoroughly.

The most significant result is the steep rise in the number of irregular/peculiar galaxies compared with their incidence in the local data. It is the increase which appears to produce the faint blue galaxy excess — $\simeq$ half the galaxy population at $I = 22$. 30% of them are either interacting or have multiple cores; these have the same steep slope as the isolated systems, confirming the tentative results of Colless *et al.* at a much higher resolution.

How reliable are these results? We note that our classification is independent of other groups in the MDS team, in particular Driver *et al.* (1994) have independently classified a smaller sample of MDS galaxies and found a similar increasing proportion of irregular/peculiar galaxies at faint magnitudes. This suggests that the result is robust with respect to observer-observer classification errors. We also wished to assess the objective reliability; our most important concern, in view of our results, was that spiral disks might be hard to identify at high-redshifts leading to misclassification as irregulars (due to spiral structure) or ellipticals (if not seen at all). Since for an object at $z = 0.7$ the $2\sigma$/pixel surface brightness limit in the HST data scales to typically $22.6\,B$ mags/arcsec$^2$ at $z = 0$ disks should still be easily visible. To check this we took a sample of CCD images of nearby galaxies and simulated observing the galaxies at $z = 0.7$ with WFPC2. A blind classification recovered the initial Hubble types; we generally found typical spiral disks were indeed very evident in the degraded data as expected. We conclude that there are no significant selection effects in our morphological classification due to the high redshift and faint magnitudes.

The physical implication for the steep slope of the optical counts is clear. At bright magnitudes ($B < 19$) there must be some incompleteness in the Schmidt plate counts of elliptical and spiral galaxies to explain the apparent fast evolution claimed by Maddox *et al.* (1990). We note increasing support for high normalisation elsewhere: Glazebrook *et al.* (1995B) find a local $\phi^* = 0.026\,h^3$ Mpc$^{-3}$ from a near-infrared selected sample; Steidel *et al.* (1994) find $\phi^* = 0.03\,h^3$ Mpc$^{-3}$ from a sample of luminous galaxies selected by their QSO absorption cross-sections. We feel that large-scale structures are unlikely to be the cause of this anomaly: it corresponds to physical scales of $\sim 170\,h^{-1}$ Mpc and typical clustering variances $\sigma^2$ are $< 0.05$ for scales $> 60h^{-1}$ Mpc (e.g. Loveday *et al.* 1992B, Peacock 1991). Much more likely are explanations due to the high surface-brightness cut inherent in plate data (e.g. McGaugh 1994, Metcalfe *et al.* 1994B) or uncertainties in the local luminosity function (e.g. Koo *et al.* 1993) or combinations thereof.

However the rapidly evolving component is clearly identified to $B = 24$ as being associated with objects of unusual morphology; and this population must contribute significantly to the large numbers of blue galaxies found at $B > 24$. Luminosity function analyses (Ellis *et al.* 1994) indicate that the evolving component at high redshift is luminous: $M_B = -20$ at $z \sim 0.5$, comparable to the elliptical/spiral populations. Evolution to lower luminosity, surface brightness or density (via mergers) must occur to remove these objects by the present day. While a self-terminating starburst (Babul & Rees 1992) is one possibility; the high incidence of multiply-peaked systems suggests that mergers play a strong role. If so this must have little effect on the properties of the regular elliptical and spiral population (c.f. Dalcanton 1993). The important discovery, from *HST*'s unique resolving power, is the unambiguous identification of the evolving component to $B = 24$. Extensive spectroscopy is required to determine the detailed nature and evolutionary path of this population.

## ACKNOWLEDGEMENTS

We wish to acknowledge many people for fruitful and interesting discussions, in particular: Stacey McGaugh, John Peacock and Mark Dickinson. We especially thank Mike Pierce and Vicki Sarajedini for their help in supplying the NGC data and other members of the Medium Deep Survey team; Stefano Casertano, John Huchra, Simon Driver and Rogier Windhorst in particular. This paper is based on observations with the NASA/ESA *Hubble Space Telescope*, obtained at the Space Telescope Science Institute, which is operated by the Association of Universities for Research in Astronomy Inc., under NASA contract NAS5-26555. Coordination and analysis of data from the Medium Deep Survey is funded by STScI grants GO2684.0X.87A and GO3917.OX.91A. U.K. data reduction and analysis was performed with computer hardware supplied by STARLINK. KGB, RSE and BXS acknowledge funding from PPARC.

## FIGURE CAPTIONS

**Figure 1.**    (a) The number-magnitude counts for all the MDS galaxies compared with the ground-based counts of Tyson (1988), Lilly *et al.* (1991) and Driver (1994). (b–d) The MDS galaxies split according to morphological classes. The lines are the no-evolution predictions described in the text for $\phi^* = 0.03\,h^3\,\mathrm{Mpc}^{-3}$ (solid) and $\phi^* = 0.015\,h^3\,\mathrm{Mpc}^{-3}$ (short dash) normalisations. The extra long dash line on (b) shows the effect of a flat faint-end luminosity function on the E/S0 prediction for the high $\phi^*$.

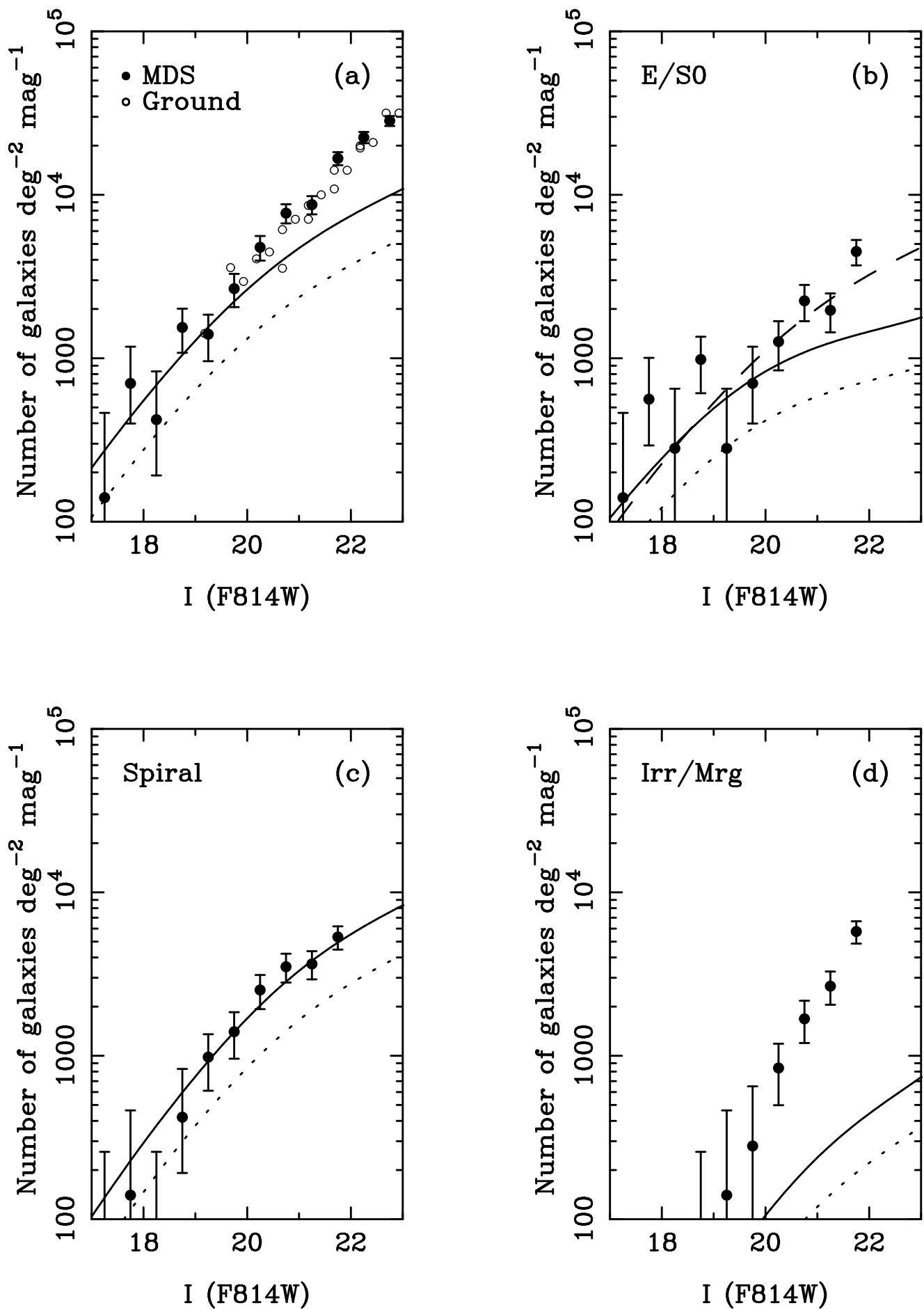

Fig. 1

This paper has been produced using the Blackwell Scientific Publications TEX macros.